\documentclass{PoS}
\usepackage{graphicx}
\def\i{{\,\scriptsize I}}

\FullConference{The 8th European VLBI Network Symposium\\
		September 26-29, 2006\\
		Toru\'n, Poland}

\title{Compact radio jets and nuclear regions in active galaxies}

\ShortTitle{Jets and nuclear regions in active galaxies}

\author{Andrei Lobanov\\
        Max-Planck-Institut f\"ur Radioastronomie, Bonn, Germany\\
        E-mail: \email{alobanov@mpifr-bonn.mpg.de}}

\abstract{VLBI observations of relativistic outflows (jets) in
galactic nuclei, complemented with detailed studies made in other
spectral domains, have become an effective tool for investigating the
physics of nuclear regions in galaxies. High-resolution radio
observations access directly the regions where the jets are formed,
and trace their evolution and interaction with the nuclear
environment. The emission properties, dynamics, and evolution of jets
in AGN are intimately connected to the characteristics of the central
supermassive black hole, accretion disk, and broad-line region (BLR) in
active galaxies.  Large VLBI surveys (15\,GHz VLBA survey, MOJAVE) and
dedicated monitoring programmes follow systematically the evolution of
several hundreds of relativistic jets.  These observations, combined
with optical and X-ray studies, yield arguably the most detailed
picture of the galactic nuclei. Recent results from studies of the
nuclear regions in several active galaxies with prominent outflows are
reviewed in this contribution.}

\begin{document}

\section{Introduction}
\label{lobanov2:sec1}

Substantial progress achieved during the past decade in studies of
active galactic nuclei (see~\cite{lobanov2006a} for a review of recent
results) has brought an increasingly wider recognition of the ubiquity
of relativistic outflows (jets) in galactic
nuclei~\cite{falcke2001,zensus1997}, turning them into an effective
probe of nuclear regions in galaxies~\cite{lobanov2005}. Emission
properties, dynamics, and evolution of an extragalactic jet are
intimately connected to the characteristics of the supermassive black
hole, accretion disk, and broad-line region (BLR) in the nucleus of the host
galaxy~\cite{lobanov2006a}.  The jet continuum emission is dominated by
non-thermal synchrotron and inverse-Compton
radiation~\cite{unwin1997}. The synchrotron mechanism plays a more
prominent role in the radio domain and the properties of the emitting
material can be assessed using the turnover point in the synchrotron
spectrum~\cite{lobanov1998b}, synchrotron
self-absorption~\cite{lobanov1998a}, and free-free absorption in the
plasma~\cite{kadler2004,walker2000}.

High-resolution radio observations access directly the regions where
the jets are formed~\cite{junor1999} and trace their evolution and
interaction with the nuclear environment~\cite{mundell2003}. Evolution
of compact radio emission from several hundreds of extragalactic
relativistic jets is now systematically studied with dedicated
monitoring programmes and large surveys using very long baseline
interferometry (VLBI) such as the 15\,GHz VLBA\footnote{Very Long Baseline
Array of National Radio Astronomy Observatory, USA}
survey~\cite{kellermann2004} and MOJAVE~\cite{lister2005}.  These
studies, combined with optical and X-ray studies, yield arguably the
most detailed picture of the galactic
nuclei~\cite{marscher2005}. Presented below is a brief summary of
recent results in this field, outlining the relation between jets,
supermassive black holes, accretion disks, and BLR in
prominent active galactic nuclei (AGN). In this respect, this review
is complementary to other recent
reviews~\cite{camenzind2005,konigl2006,marscher2005} focused on
formation and propagation of extragalactic relativistic jets.

\section{Anatomy of jets}

\begin{figure}[t]
\centering
\includegraphics[width=0.48\textwidth,angle=-90,bb=125 80 472 761,clip=true]{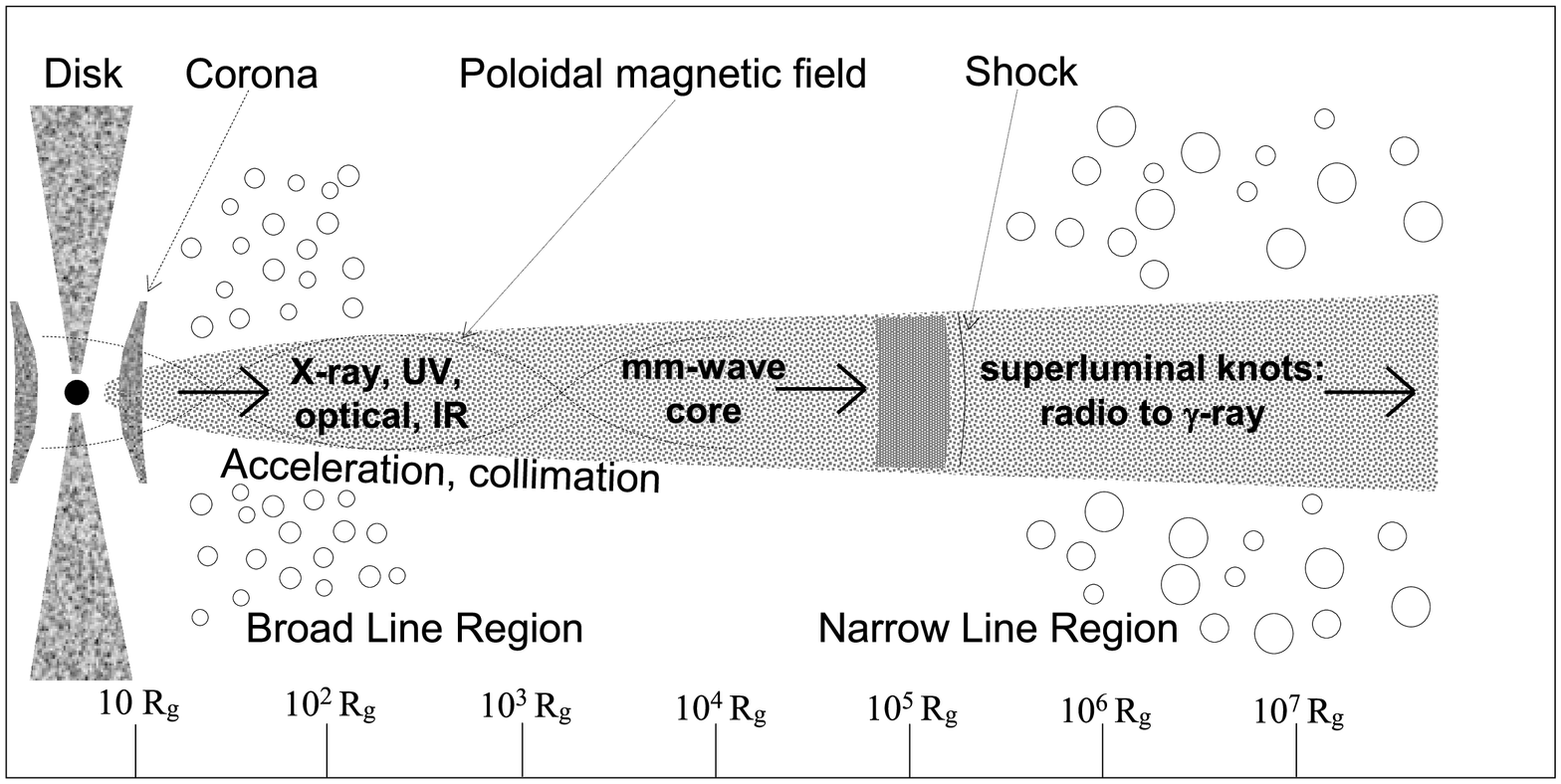}
\caption{Basic sketch of a radio-loud AGN featuring a relativistic jet and
its environment. The scale is logarithmic beyond 10\,$R_\mathrm{g}$ and vertical dimensions are not drawn to scale.
Regions of the jet mainly contributing to various parts of the broadband 
spectrum are indicated. Adapted from~\cite{marscher2005}.
}
\label{lobanov:fg01}
\end{figure}

Jets in active galaxies are formed in the immediate vicinity of the
central black hole~\cite{camenzind2005} and they interact with every
major constituent of AGN (see Figure~\ref{lobanov:fg01} and
Table~\ref{lobanov2:tb1}). The jets carry away a fraction of the
angular momentum and energy stored in the accretion
flow~\cite{hujeirat2003} or corona (in low luminosity
AGN~\cite{merloni2002}) and in the rotation of the central black
hole~\cite{koide2002,komissarov2005,semenov2004}.

\begin{table}[h]
\caption{Characteristic scales in the nuclear regions in active galaxies}
\label{lobanov2:tb1}
\small
\begin{center}
\begin{tabular}{rccccc}\hline\hline
   & $l$ & $l_8$ & $\theta_\mathrm{Gpc}$ & $\tau_c$ & $\tau_\mathrm{orb}$ \\ 
  & [$R_\mathrm{g}$] & [pc] & [mas]& [yr] & [yr] \\ \hline
Event horizon:           &1--2          &$10^{-5}$           &$5\times 10^{-6}$    &0.0001     & 0.001 \\
Ergosphere:              &1--2          &$10^{-5}$           &$5\times 10^{-6}$    &0.0001     & 0.001 \\
Corona:                  &10$^1$--10$^2$&$10^{-4}$--$10^{-3}$&$5\times 10^{-4}$    &0.001--0.01& 0.2--0.5 \\
Accretion disk:          &10$^1$--10$^3$&$10^{-4}$--$10^{-2}$&$0.005$              &0.001--0.1 & 0.2--15 \\
{\bf Jet formation:}          &$>$10$^2$     &$>$$10^{-3}$        &$>$$5\times 10^{-4}$ &$>$0.01    & $>$0.5 \\
{\bf Jet visible in the radio:}&$>$10$^3$     &$>$$10^{-2}$        &$>$$0.005$           &$>$0.1     & $>$15 \\ 
Broad-line region:       &10$^2$--10$^5$&$10^{-3}$--1        &$0.05$               &0.01--10   & 0.5--15000 \\
Molecular torus:         &$>$10$^5$     &$>$1                &$>$$0.5$             &$>$10      & $>$15000 \\
Narrow-line region:      &$>$10$^6$     &$>$10               &$>$5                 &$>$100     & $>$500000 \\ \hline
\end{tabular}
\end{center}
{Column designation:}~$l$ -- dimensionless scale in units of the
gravitational radius, $G\,M/c^2$; $l_8$ -- corresponding linear scale,
for a black hole with a mass of $5\times 10^8\,M_{\odot}$;
$\theta_\mathrm{Gpc}$ -- corresponding largest angular scale at 1\,Gpc
distance; $\tau_c$ -- rest frame light crossing time;
$\tau_\mathrm{orb}$ -- rest frame orbital period, for a circular
Keplerian orbit. Adapted from~\cite{lobanov2006a}
\end{table}

The production of highly-relativistic outflows requires a large
fraction of the energy to be converted to Poynting flux in the very
central region~\cite{sikora2005}.  The efficiency of this process may
depend on the spin of the central black hole~\cite{meier1999}. The
collimation of such a jet requires either a large-scale poloidal
magnetic field threading the disk~\cite{spruit1997} or a slower and
more massive MHD outflow launched at larger disk radii by centrifugal
forces~\cite{bogovalov2005,tsinganos2002}. A two-zone version of such
a hybrid outflow model is known as the ``two-fluid
model''~\cite{sol1989}.

At distances of $\sim 10^3\,R_\mathrm{g}$ ($R_\mathrm{g} = G\,M/c^2$ is the gravitational radius of a black hole), the jets become visible in
the radio regime, which makes high-resolution VLBI observations a tool
of choice for probing directly the physical conditions in AGN on such
small scales~\cite{junor1999,krichbaum2004}. Recent studies indicate
that at distances of $10^3$--$10^5\,R_\mathrm{g}$ ($\lesssim 1$\,pc)
the jets are likely to be dominated by pure electromagnetic processes
such as Poynting flux~\cite{sikora2005} or have both MHD and
electrodynamic components~\cite{meier2003}. The flowing plasma is
likely to be dominated by electron-positron
pairs~\cite{wardle1998,hirotani2005}, although a dynamically
significant proton component cannot be completely ruled out at the
moment~\cite{celotti1993}.

The magnetic field is
believed to play an important role in accelerating and collimating
extragalactic jets on parsec scales~\cite{vlahakis2004}. How far the
magnetic field dominated region extends in extragalactic jet is still
a matter of debate~\cite{sikora2005}.  Nevertheless, it is possible
to identify three distinct regions with different physical mechanisms
dominating the observed properties of the jets: {\em 1)~ultracompact jets}
where collimation and acceleration of the flow occurs, {\em 2)~parsec-scale
flows} dominated by relativistic shocks, and {\em 3)~large-scale jets} where
plasma instabilities dominate the flow.

\begin{figure}[h]
\includegraphics[width=0.99\textwidth,angle=0,bb=0 0 1129 499,clip=true]{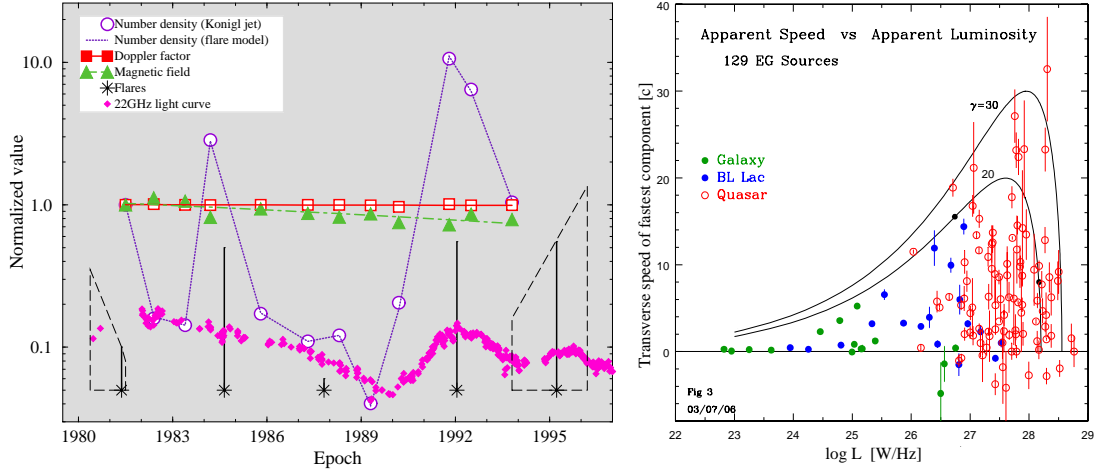}
\caption{ {\em Left panel:}~Relative changes of the Doppler factor and
magnetic field in the VLBI core in 3C\,345, obtained by applying
K\"onigl jet model to the measured flux density and frequency of the
synchrotron peak in the spectrum. All quantities are normalized to
their respective values at the first epoch, $t_0=1981.5$. Open circles
denote the particle density required for maintaining a constant
Doppler factor. The dotted line shows the particle density, as
represented by five exponential flares. The resulting 
Doppler factor (open squares) and magnetic field (filled triangles)
are also shown, with lines representing linear fits to the respective
quantity. The light curve of the total flux density at 22\,GHz is plotted for
comparison, scaled down by a factor of 100. Reproduced
from~\cite{lobanov1999}.  {\em Right panel:}~Values of apparent
transverse speed, $\beta_\mathrm{app}$, and apparent luminosity, $L$
plotted for the fastest component of 129 radio sources in the
15\,GHz VLBA survey~\cite{kellermann2004}. Colours indicate different
host galaxy types. The aspect curves are the loci of
($\beta_\mathrm{app},L$) for sources with Lorentz factors
$\Gamma_\mathrm{j} = 30$ and 20, and $L_0 = 1\times
10^{25}$\,W\,Hz$^{-1}$.  Reproduced from~\cite{cohen2006}.  }
\label{lobanov:fg02}
\end{figure}

\subsection{Ultracompact jets}

Ultracompact jets observed down to sub-parsec scales typically show
strongly variable but weakly polarized emission (possibly due to
limited resolution of the observations).  Compelling evidence exists
for acceleration~\cite{bach2005} and
collimation~\cite{junor1999,krichbaum2004,krichbaum2006} of the flows
on these scales, which is most likely driven by the magnetic
field~\cite{vlahakis2004}.  The ultracompact outflows are probably
dominated by electromagnetic processes~\cite{meier2003,sikora2005}
and they become visible in the radio regime (identified as compact
``cores'' of jets ) at the point where the jet becomes optically thin
for radio emission~\cite{lobanov1998a,lobanov1999}. The ultracompact
jets do not appear to have strong shocks~\cite{lobanov1998b} and their
basic properties are successfully described by quasi-stationary
flows~\cite{konigl1981}.  The evolution and variability of
ultra-compact jets (Figure~\ref{lobanov:fg02}, left panel) can be
explained by smooth changes in particle density of the flowing plasma,
associated with the nuclear flares in the central
engine~\cite{lobanov1999}. Intrinsic brightness temperatures of the
ultracompact jets are estimated to be (1--$5) \times
10^{11}$\,K~\cite{lobanov2000}, implying that the energy losses are
dominated by the inverse-Compton process~\cite{kellermann1969}.

Quasi-periodic variability of the radio emission from the ultracompact
jets is most likely related to instabilities and non-stationary
processes in the accretion disks around central black holes in
AGN~\cite{igumenschev1999,lobanov2005b}. Alternative explanations
involve binary black hole systems in which flares are caused by
passages of the secondary through the accretion disk around the
primary~\cite{ivanov1998,lehto1996}. These models, however, require
very tight binary systems, with orbits of the secondary lying well
within $10^3a$ Schwarzschild radii of the primary (between 20 and 100
Schwarzschild radii, in the celebrated case of
OJ\,287~\cite{lehto1996}). This poses inevitable problems for
maintaining an accretion disk around the primary (for massive
secondaries~\cite{lobanov2006}) or rapid alignment of the secondary
with the plane of the accretion disk (for small
secondaries~\cite{ivanov1999}). Discrepancy between the predicted and
actual epoch of the latest outburst in OJ\,287~\cite{valtonen2006}
indicates further that the observed behaviour is not easy to be
reproduced by a binary black hole scenario and it is indeed more
likely to result from a quasi-periodic process in the disk, similarly
to the flaring activity observed in 3C\,345~\cite{lobanov2005b}.

\begin{figure}[t]
\centering \includegraphics[width=0.98\textwidth,angle=0,bb=0 0 1632
735,clip=true]{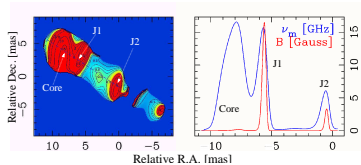}
\caption{{\em Left panel:}~Distribution of the synchrotron turnover
frequency, $\nu_\mathrm{m}$, in the jet in 3C\,273. The distribution
is obtained from multifrequency VLBI observations and shows three
regions of increased $\nu_\mathrm{m}$: one is in the ``core'' of the
jet and the other two are coincident with two bright superluminal
features in the jet (denoted J1 and J2).  {\em Right panel:}~Profiles
of the turnover frequency (blue) and magnetic field strength (red)
taken in the jet, along the dashed line shown in the left panel. The
low magnetic field in the core region indicates that the field is
tangled or intrinsically weak there. The magnetic field profile shows
strong spikes at the locations of the features J1 and J2, implying
that these are most likely due to magnetic field compression in
strong shocks propagating in the jet. Adapted
from~\cite{lobanov1997}.}
\label{lobanov:fg03}
\end{figure}

\subsection{Parsec-scale flows: shocks and instabilities}

Parsec-scale outflows are characterized by pronounced curvature of
trajectories of superluminal
components~\cite{kellermann2004,lobanov1999}, rapid changes of
velocity and flux density, and predominantly transverse magnetic
field~\cite{jorstad2005}. Statistical studies of speed and brightness
temperature distributions observed in the superluminal features
propagating on parsec scales indicate that the jet population has an
envelope Lorentz factor of $\approx 30$ and an unbeamed luminosity of
$\sim 1\times 10^{25}$\,W\,Hz$^{-1}$~\cite{cohen2006} (right panel
of Figure~\ref{lobanov:fg02}), with the brightness temperature of the
emitting plasma reaching $\sim 5\times 10^{10}$\,K~\cite{lobanov2000},
which is close to the equipartition limit~\cite{readhead1994}.

Relativistic shocks are expected to be prominent on these scales,
which is manifested by strong polarization~\cite{ros2000} and rapid
evolution of the turnover frequency of synchrotron
emission~\cite{lobanov1997}.  Mapping the turnover frequency
distribution provides also a sensitive diagnostic of shocks and plasma
instabilities in relativistic flows~\cite{lobanov1998b}.  Shocks are
particularly evident in the profiles of magnetic field obtained from
the turnover frequency images (see Figure~\ref{lobanov:fg03}).

\begin{figure}
\centering
\includegraphics[width=0.99\textwidth,angle=0,bb=0 0 1227 502,clip=true]{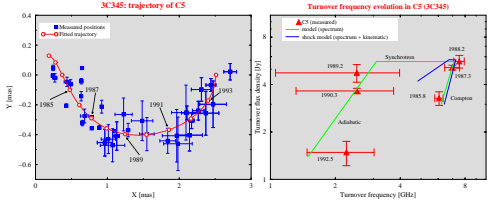}
\caption{{\em Left panel:}~The observed trajectory of C5 in the plane of the sky. The 
dotted line represents the combined polynomial fits to the component's
$x$ and $y$ offsets from the core. Open circles mark the locations on
the trajectory which are equally spaced in time at an interval of 1
year. {\em Right panel:}~Evolution of the synchrotron turnover changes in  C5. The dotted line shows an evolutionary track consistent
with different stages of the evolution of the shock (Compton-loss, synchrotron-loss, and adiabatic-loss stage).  The dot--dashed line shows how the
original fit for C5 must be changed to satisfy also the observed
trajectory of the component (shown in the left panel). The shock description 
can only be applied at early stages of the component evolution, indicating that the shock should dissipate rapidly, on scales of $\lesssim 100$\,pc. Adapted from~\cite{lobanov1999}.
}
\label{lobanov:fg04}
\end{figure}

Complex evolution of shocked regions is revealed in
observations~\cite{gomez2001,jorstad2005,lobanov1999} and numerical
simulations~\cite{agudo2001} of parsec-scale outflows. However, the
shocks are shown to dissipate rapidly~\cite{lobanov1999} and
shock-dominated regions are not likely to extend beyond $\sim
100$\,pc. This can be exemplified by the kinematic and spectral
evolution observed in the jet component C5 in 3C\,345
(Figure~\ref{lobanov:fg04}). The shock model can reproduce the
observed spectral and kinematic changes only during the initial stages
of the component evolution. This implies that strong shocks should
dissipate rapidly in the jets, on scales of $\lesssim
100$\,pc~\cite{lobanov1999}.

Observed morphologies of parsec-scale jets and trajectories of
superluminal features propagating in the jets are often described in
terms of a helical geometry~\cite{steffen1995}, with helicity arising
from some periodic process in the nucleus. Jet precession, both in
single~\cite{lu2005} and binary~\cite{caproni2004} black hole systems,
have been commonly sought to be responsible for the observed helicity
on parsec scales, but it may lead to rather non-physical parameters of
the nuclear region~(see \cite{lobanov2005b} for discussion). It is
more likely that the jet precession in AGN acts on time scales of
$\gtrsim 10^{4}$ years and manifests itself in kiloparsec-scale
structures~\cite{gower1982} where it also becomes affected by the
motion of the host galaxy~\cite{zaninetti1989}. Determining the
precession parameters from observations of parsec-scale structures
requires detailed description of relativistic effects and source
geometry. This yields much longer orbital and precession
periods~\cite{lobanov2005b} than those obtained from associating
quasi-periodic changes observed on these scales directly with
precession~\cite{caproni2004}. On parsec scales, the observed helical
patterns may be produced not exclusively by precession but also by
other processes, including Kelvin-Helmholtz
instability~\cite{hardee2003} and rotation of the
flow~\cite{camenzind1992}. The flow rotation is particularly relevant
for explaining periodic changes of the ``ejection angle'' (flow
direction in the immediate vicinity of the VLBI core on
sub-milliarcsecond scales) reported in several prominent
objects~\cite{abraham1998,abraham1999,mutel2005}.

\subsection{Large-scale jets}

\begin{figure}[t]
\centering
\includegraphics[totalheight=0.95\textwidth,angle=-90,bb=118 86 472 754,clip=true]{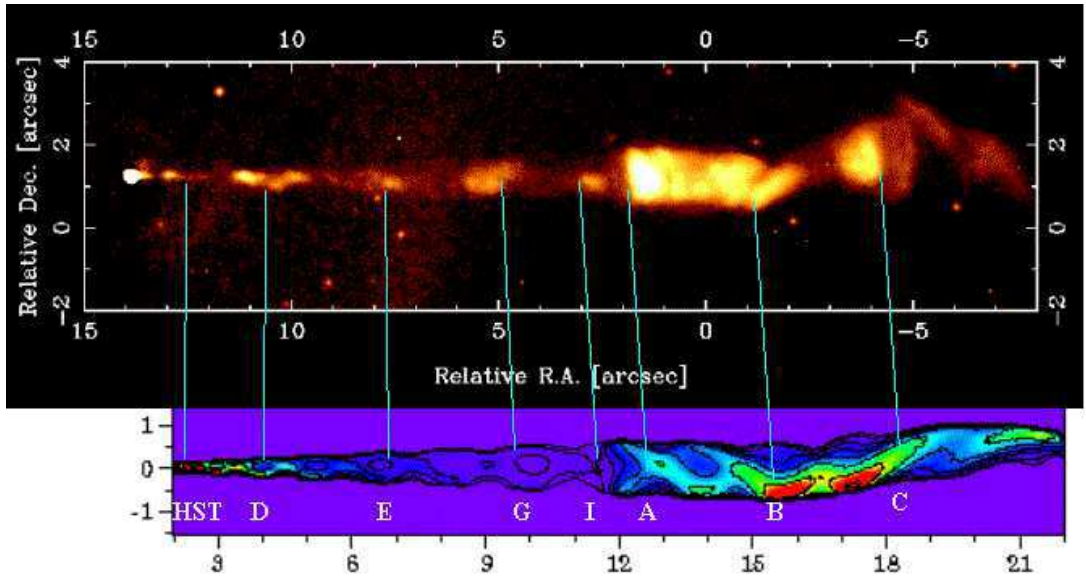}
\caption{Top: HST image of the jet in M\,87. Bottom: Model based on
linear description of Kelvin-Helmholtz instability developing in the
jet. Line-of-sight synchrotron intensity image and contours at a jet
viewing angle of $40^\circ$ including all light travel time effects
and time delays. The beginning point is about $2^{\prime\prime}$ out
from the nucleus with the end point at about $22^{\prime\prime}$. The
spatial twist is such that the twisted flow is more towards the
observer at the bottom of the jet and is more away at the top of the
jet.  Thus, knots A and C would be Doppler de-boosted and knot B would
be Doppler boosted. Reproduced from~\cite{lobanov2003}.  }
\label{lobanov:fg05}
\end{figure}

On scales larger than $\sim 100$\,pc, instabilities (most importantly,
Kelvin-Helmholtz instability) determine at large the observed
structure and dynamics of extragalactic
jets~\cite{lobanov2001,lobanov2003,perucho2006}.  The elliptical mode
of the instability is responsible for appearance of thread-like
features in the jet interior, while overall oscillations of the jet
ridge line are caused by the helical surface mode. Successful attempts
have been made to represent the observed brightness distribution of
radio emission on these scales, using linear perturbation theory of
Kelvin-Helmholtz instability~\cite{lobanov2003} and a spine-sheath
(analogous to the two-fluid) description of relativistic
flows~\cite{canvin2005,laing2004}. Figure~\ref{lobanov:fg05}
illustrates application of linear Kelvin-Helmholtz model to
reproducing the radio brightness distribution in the jet of
M87~\cite{lobanov2003}. Non-linear evolution of Kelvin-Helmholtz
instability~\cite{perucho2004a,perucho2004b} and stratification of the
flow~\cite{perucho2005} play important roles in large-scale jets.
Similarly to stellar jets, rotation of the flow is expected to be
important for extragalactic jets~\cite{fendt1997}, but observational
evidence remains very limited on this issue.

\section{Jets and nuclear regions in AGN}

A number of recent studies have used jets to probe physical conditions
in the central regions of AGN.  These studies have probed physical 
conditions in the compact relativistic flows, properties of atomic and molecular material in circumnuclear regions of AGN, and connection between relativistic outflows, accretion disks, and BLRs.

\begin{figure}[t]
\centering
\includegraphics[width=0.48\textwidth,angle=0,bb=12 12 576 480,clip=true]{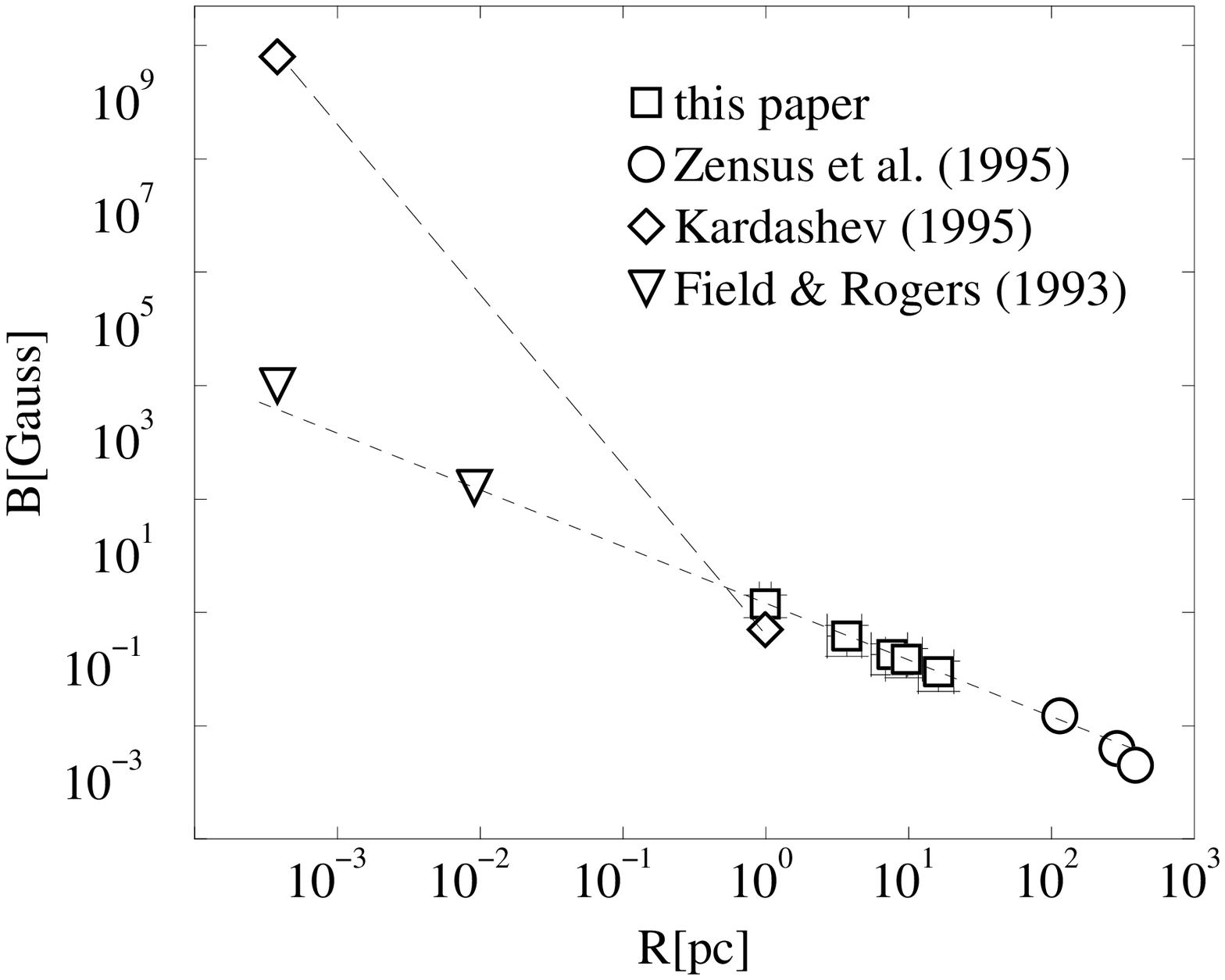}
\includegraphics[width=0.48\textwidth,angle=0,bb=12 12 576 480,clip=true]{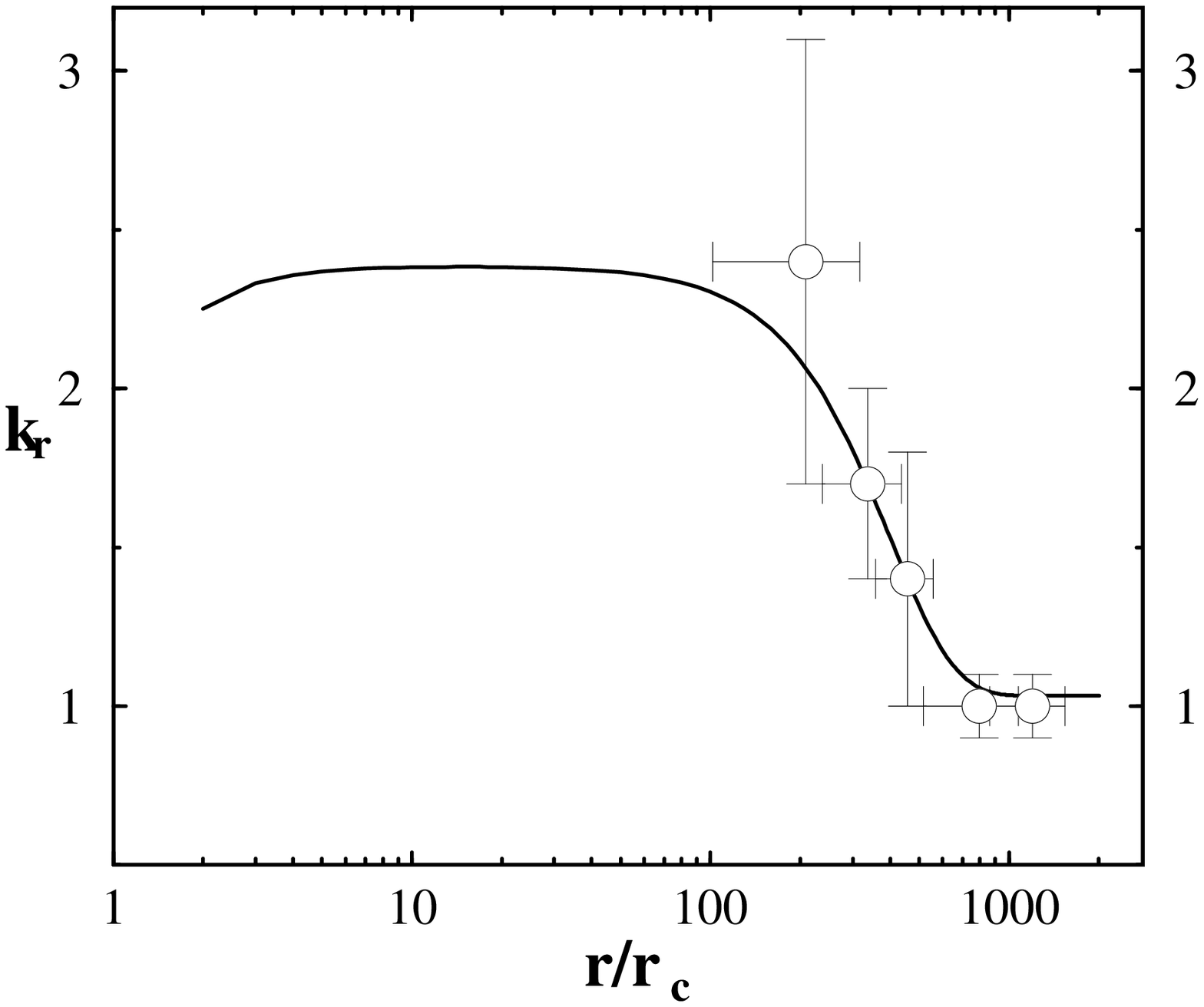}
\caption{{\em Left panel:}~Magnetic field distribution in the jet in
3C\,345. Squares show the magnetic field in the compact jet derived
from the frequency-dependent shift of the core. Circles are the
homogeneous synchrotron model estimates of magnetic field in the
extended jet components.  Triangles show the
characteristic magnetic field values from a model of magnetized
accretion disk. Diamonds are the theoretical
estimates for the dipole magnetic field
around a supermassive rotating black hole. {\em Right panel:}~Opacity
in the jet in 3C\,309.1. Circles are the measured values of the core
shift power index $k_\mathrm{r}$ in 3C\,309.1 at different
frequencies; solid line shows changes of $k_\mathrm{r}$ due to
pressure gradients in the BLR clouds supported by thermal pressure and
maintaining a mass distribution with spherically symmetrical
gravitational potential. The cloud region extends up to
400\,$r_\mathrm{s}$ ($r_\mathrm{s}$ refers to the distance at which
the jet becomes supersonic). The equipartition regime is approached at
the outer boundary of the cloud region, with $k_\mathrm{r} =
1$. Significant deviations from the equipartition are seen on smaller
scales, resulting in stronger self-absorption in the inner parts of
the jet. Reproduced from~\cite{lobanov1998a}.  }
\label{lobanov:fg06}
\end{figure}

\subsection{Jets and nuclear regions of AGN}

Synchrotron self-absorption and external absorption in the
ultracompact jets (VLBI ``cores'') can be used effectively for
determining the properties of the flow itself and its
environment~\cite{lobanov1998a}.  Absolute position of the core,
$r_\mathrm{c}$, varies with the observing frequency, $\nu$, so that
$r_\mathrm{c} \propto \nu^{-1/k_\mathrm{r}}$,~\cite{konigl1981}. If
the core is self-absorbed and in equipartition, the power index
$k_\mathrm{r}=1$~\cite{blandford1979}.  Changes of the core position
measured between three or more frequencies can be used for determining
the value of $k_\mathrm{r}$ and estimating the strength of the
magnetic field, $B_\mathrm{core}$, in the nuclear region and the
offset, $R_\mathrm{core}$, of the observed core positions from the
true base of the jet (see Figure~\ref{lobanov:fg06}; left panel).  The
combination of these gives an estimate for the mass of the central black
hole $M_\mathrm{bh} \approx 7\times 10^9\,M_\odot\,
(B_\mathrm{core}/\mathrm{G})^{1/2} (R_\mathrm{core}/\mathrm{pc})^{3/2}$.

Core shift measurements provide estimates of the total (kinetic +
magnetic field) power, the synchrotron luminosity, and the maximum
brightness temperature, $T_\mathrm{b,max}$ in the jets. The ratio of
particle energy to magnetic field energy can also be estimated, from
the derived $T_\mathrm{b,max}$.  This enables testing the original
K\"onigl model~\cite{konigl1981} and several of its later
modifications (e.g., \cite{hutter1986,bloom1996}).  The known distance
from the nucleus to the jet origin will also enable constraining the
self-similar jet model~\cite{marscher1995} and the particle-cascade
model~\cite{blandford1995}.

Recent studies of free-free absorption in AGN indicate the presence of
dense, ionized circumnuclear material with $T_\mathrm{e} \approx
10^4$\,K distributed within a fraction of a parsec of the central
nucleus~\cite{lobanov1998a,walker2000}.  Properties of the
circumnuclear material can also be studied using the variability of
the power index $k_\mathrm{r}$ with frequency. This variability
results from pressure and density gradients or absorption in the
surrounding medium, most likely associated with BLR. Changes of
$k_\mathrm{r}$ with frequency, if measured with required precision,
can be used to estimate the size, particle density, and temperature of
the absorbing material surrounding the jets (see
Figure~\ref{lobanov:fg06}; right panel). Estimates of the black hole
mass and the BLR size obtained from the core-shift measurements can
be compared with the respective estimates obtained from the
reverberation mapping and applications of the
$M_\mathrm{bh}$--$\sigma_\star$ relation.

\subsection{Atomic and molecular absorption}

Opacity and absorption in the nuclear regions of AGN have been probed
effectively using the non-thermal continuum emission as a background
source~\cite{lobanov2005}. Absorption due to several atomic and
molecular species (most notably due to H\i, CO, OH, and HCO$^+$) has
been detected in a number of extragalactic objects. OH absorption has
been used to probe the conditions in warm neutral
gas~\cite{goikoechea2004,kloeckner2005}, and CO and H\i\ absorption
have been used to study the molecular
tori~\cite{conway1999,pedlar2004} at a linear resolution often smaller
than a parsec~\cite{mundell2003}. These observations have revealed the
presence of neutral gas in a molecular torus in NGC\,4151 and in a
rotating outflow surrounding the relativistic jet in
1946$+$708~\cite{peck2001}.


\subsection{Jet-disk and jet-BLR connections}

Connection between accretion disks and relativistic
outflows~\cite{hujeirat2003} has been explored using correlations
between variability of X-ray emission produced in the inner regions of
accretion disks and ejections of relativistic plasma into the
flow~\cite{marscher2002}. The jets can also play a role in the
generation of broad emission lines in AGN. The beamed continuum
emission from relativistic jet plasma can illuminate atomic material
moving in a sub-relativistic outflow from the nucleus, producing broad
line emission in a conically shaped region located at a significant
distance above the accretion disk~\cite{arshakian2006}. Magnetically
confined outflows can also contain information about the dynamic
evolution of the central engine, for instance that of a binary black
hole system~\cite{lobanov2005b}. This approach can be used for
explaining, within a single framework, the observed optical
variability, kinematics, and flux density changes of superluminal
features embedded in radio jets.

\section{Conclusion}

Extragalactic jets are an excellent laboratory for studying physics of
relativistic outflows and probing conditions in the central regions of
active galaxies. Recent studies of extragalactic jets show that they
are formed in the immediate vicinity of central black holes in
galaxies and carry away a substantial fraction of the angular momentum
and energy stored in the accretion flow and rotation of the black
hole. The jets are most likely collimated and accelerated by
electromagnetic fields. Relativistic shocks are present in the flows
on small scales, but dissipate on scales of $\lesssim 100$\,pc. Plasma
instabilities dominate the flows on larger scales.  Convincing
observational evidence exists, connecting ejections of material into
the flow with instabilities in the inner accretion disks. In
radio-loud objects, continuum emission from the jets may also drive
broad emission lines generated in sub-relativistic outflows
surrounding the jets.  Magnetically confined outflows may preserve
information about the dynamics state of the central region, allowing
detailed investigations of jet precession and binary black hole
evolution to be made. This makes studies of extragalactic jets a
powerful tool for addressing the general questions of physics and
evolution of nuclear activity in galaxies.



\begin{thebibliography}{99}

\bibitem{abraham1998} Z. Abraham, E.A. Carrara: ApJ \textbf{496}, 172 (1998)

\bibitem{abraham1999} Z. Abraham, G.E. Romero: A\&A \textbf{344}, 61 (1999)

\bibitem{agudo2001} I. Agudo, J.L. G\'omez, J.M. Mart\'i, et al.: ApJ \textbf{549}, 183 (2001)

\bibitem{arshakian2006} T.G. Arshakian, A.P. Lobanov, V.H. Chavushyan, et al.: A\&A \textit{subm.} (2006), \texttt{astro-ph/0512393}

\bibitem{bach2005} U. Bach, M. Kadler, T.P. Krichbaum, et al.: Multi-Frequency and Multi-Epoch VLBI Study of Cygnus A. In: \textit{Future Directions in High Resolution Astronomy: The 10th Anniversary of the VLBA}, ASP Conference Proceedings, Vol. 340, ed. by J. Romney, M. Reed (ASP, San Fransisco 2005) pp. 30--34

\bibitem{blandford1979} R.D. Blandford, A. K\"onigl: ApJ \textbf{232},
34 (1979)

\bibitem{blandford1995} R.D. Blandford, A. Levinson: ApJ \textbf{441},
79 (1995)

\bibitem{bloom1996} S.D. Bloom, A.P. Marscher: ApJ \textbf{461}, 657
(1996)

\bibitem{bogovalov2005} S.V. Bogovalov, K. Tsinganos: MNRAS
\textbf{357}, 918 (2005)

\bibitem{camenzind2005} M. Camenzind: MemSAIt \textbf{76}, 98 (2005)

\bibitem{camenzind1992} M. Camenzind, M. Krockenberger: A\&A
\textbf{255}, 59 (1992)

\bibitem{canvin2005} J.R. Canvin, R.A. Laing, A.H. Bridle,
W.D. Cotton: MNRAS \textbf{363}, 1223 (2005)

\bibitem{caproni2004} A. Caproni, Z. Abraham: ApJ \textbf{602}, 625 (2004)

\bibitem{celotti1993} A. Celotti, A.C. Fabian: MNRAS \textbf{264}, 228 (1993)

\bibitem{conway1999} J.E. Conway: New Astron. Rev. \textbf{43}, 509 (1999)

\bibitem{cohen2006} M.H. Cohen, M.Lister, D. Homan, et al.: ApJ, {\em
subm.} (2006)

\bibitem{falcke2001} H. Falcke: Rev. Mod. Astron. \textbf{14}, 15 (2001)

\bibitem{fendt1997} C. Fendt: A\&A \textbf{323}, 999 (1997)

\bibitem{gomez2001} J.L. G\'omez, A.P. Marscher, A. Alberdi, et al.:
ApJ \textbf{561}, 161 (2001)

\bibitem{goikoechea2004} J.R. Goikoechea, J. Mart\'in-Piintado, J. 
Chernicharo: ApJ \textbf{619}, 291 (2005)

\bibitem{gower1982} A.C. Gower, P.C. Gregory, W.G. Unruh,
J.B. Hutchings: ApJ \textbf{262}, 478 (1982)

\bibitem{hardee2003} P.E. Hardee: ApJ \textbf{597}, 798 (2003)

\bibitem{hirotani2005} K. Hirotani: ApJ \textbf{619}, 73 (2005)

\bibitem{hujeirat2003} M. Hujeirat, M. Livio, M. Camenzind, et al.:
A\&A \textbf{408}, 415 (2003)

\bibitem{hutter1986} D.J. Hutter, S.L. Mufson: ApJ \textbf{301}, 50 (1986)

\bibitem{igumenschev1999} I.V. Igumenschev, M.A. Abramowicz: MNRAS
\textbf{303}, 309 (1999)

\bibitem{ivanov1998} P.B. Ivanov, I.V. Igumenschev, I.D. Novikov:
ApJ \textbf{507} 131 (1998)

\bibitem{ivanov1999} P.B. Ivanov, J.C.B. Papaloizou, A.G. Polnarev:
MNRAS \textbf{307}, 79 (1999)

\bibitem{jorstad2005} S.G. Jorstad, A.P. Marscher, M.L. Lister, et al.:
AJ \textbf{130}, 1418 (2005)

\bibitem{junor1999} W. Junor, J.A. Biretta, M. Livio: Nature \textbf{401}, 891 (1999)

\bibitem{kadler2004} M. Kadler, E. Ros, A.P. Lobanov, et al.: A\&A
\textbf{426}, 481 (2004)

\bibitem{kellermann1969} K.I. Kellermann, I.I.K. Pauliny-Toth: ApJ \textbf{155}, 71 (1969)

\bibitem{kellermann2004} K.I. Kellermann, M.L. Lister, D.C. Homan, et al.: AJ,
\textbf{609}, 539 (2004)

\bibitem{kloeckner2005} H.R. Kl\"ockner, W.A. Baan: ApSS \textbf{295},
277 (2005)

\bibitem{konigl1981} A. K\"onigl: ApJ \textbf{243}, 700 (1981)

\bibitem{konigl2006} A. K\"onigl: MemSAIt \textbf{77}, 598 (2006)

\bibitem{koide2002} S. Koide, K. Shibata, T. Kudoh, et al.: Science
\textbf{295}, 1688 (2002)

\bibitem{komissarov2005} S.S. Komissarov: MNRAS \textbf{359}, 801 (2005)


\bibitem{krichbaum2004} T.P. Krichbaum, D.A. Graham, A. Kraus, et al.:
Towards the Event Horizon -- The Vicinity of AGN at Micro-Arcsecond Resolution.
In: \textit{Proceedings of the 7th Symposium of the European VLBI Network},
ed. by R. Bachiller, F. Colomer, J.-F. Desmurs, P. de Vicente (Observatorio Astronomico Nacional, Madrid 2004) pp. 15--18

\bibitem{krichbaum2006} T.P. Krichbaum: {\em this volume}, (2006)

\bibitem{laing2004} R.A. Laing, A.H. Bridle: MNRAS \textbf{348}, 1459 (2004)

\bibitem{lehto1996} H.J. Lehto, M.J. Valtonen: ApJ \textbf{460}, 207 (1996)

\bibitem{lister2005} M.L. Lister, D.C. Homan: AJ \textbf{130}, 1389 (2005)

\bibitem{lobanov1998a} A.P. Lobanov: A\&A \textbf{390}, 79 (1998)

\bibitem{lobanov1998b} A.P. Lobanov: A\&ASS \textbf{132}, 261 (1998)

\bibitem{lobanov2005} A.P. Lobanov: MemSAItS \textbf{7}, 12 (2005)

\bibitem{lobanov2006} A.P. Lobanov: Nuclear activity in galaxies driven by supermassive binary black holes. In: \textit{Relativistic Astrophysics and Cosmology}, ESO Astrophysical Symposia, eds. B. Aschenbach, V. Burwitz, G. Hasinger, B. Leibundgut (Springer: Heidelberg 2006) {\tt astro-ph/0606198}

\bibitem{lobanov2005b} A.P. Lobanov, J. Roland: A\&A \textbf{431}, 831
(2005)

\bibitem{lobanov1999} A.P. Lobanov, J.A. Zensus: ApJ \textbf{521}, 509 (1999)

\bibitem{lobanov2001} A.P. Lobanov, J.A. Zensus: Science \textbf{284},
291 (2001)

\bibitem{lobanov2006a} A.P. Lobanov, J.A. Zensus: Active Galactic Nuclei at the Crossroads of Astrophysics. In: \textit{Exploring the Cosmic Frontier: Astrophysical Instruments for the 21$^\mathrm{st}$ Century}, ESO Astrophysical Symp. Series, ed. by A.P. Lobanov, J.A. Zensus, C. Cesarsky, P.J. Diamond (Springer, Heidelberg 2006) pp. 147--162

\bibitem{lobanov1997} A.P. Lobanov, E. Carrara, J.A. Zensus: 
Vistas in Astronomy \textbf{41}, 253 (1997)

\bibitem{lobanov2003} A.P. Lobanov, P.E. Hardee, J.A. Eilek: New Astron. Rev. \textbf{47}, 629 (2003)

\bibitem{lobanov2000} A.P. Lobanov, T.P. Krichbaum, D.A. Graham, et al.: A\&A \textbf{364}, 391 (2000)

\bibitem{lu2005} J.-F. Lu, B.-Y. Zhou: ApJ \textbf{635}, L17 (2005)

\bibitem{marscher1995} A.P. Marscher: PNAS \textbf{92}, 11439 (1995)

\bibitem{marscher2002} A.P. Marscher, S.G. Jorstad, J.L. G\'omez, et al.:
Nature \textbf{417}, 625 (2002)

\bibitem{marscher2005} A.P. Marscher: MemSAIt \textbf{76}, 13 (2005)

\bibitem{meier1999} D.L. Meier: ApJ \textbf{522}, 753 (1999)

\bibitem{meier2003} D.L. Meier: New Astron. Rev. \textbf{47}, 667 (2003)

\bibitem{merloni2002} A. Merloni, A.C. Fabian: MNRAS \textbf{332}, 165 (2002)

\bibitem{mundell2003} C.G. Mundell, J.M. Wrobel, A. Pedlar, et al.:
ApJ \textbf{583}, 192 (2003)

\bibitem{mutel2005} R.L. Mutel, G.R. Denn: ApJ \textbf{623}, 79 (2005)

\bibitem{peck2001} A.B. Peck, G.B. Taylor: ApJ \textbf{554}, 147 (2001)

\bibitem{pedlar2004} A. Pedlar: ApSS \textbf{295}, 161 (2004)

\bibitem{perucho2004a} M. Perucho, M. Hanasz, J.M. Mart\'i, et al.: A\&A \textbf{427}, 415 (2004)

\bibitem{perucho2004b} M. Perucho, J.M. Mart\'i, M. Hanasz: A\&A \textbf{427}, 431 (2004)

\bibitem{perucho2005} M. Perucho,  J.M. Mart\'i, M. Hanasz: A\&A \textbf{443}, 863 (2005)

\bibitem{perucho2006} M. Perucho, A.P. Lobanov, J.M. Mart\'i: \textit{these proceedings} (2006)

\bibitem{readhead1994} A.C.S. Readhead: ApJ \textbf{426}, 51 (1994)

\bibitem{ros2000} E. Ros, J.A. Zensus, A.P. Lobanov: A\&A
\textbf{354}, 55 (2000)

\bibitem{semenov2004} V. Semenov, S. Dyadechkin, B. Punsly: Science \textbf{305}, 978 (2004)

\bibitem{sikora2005} M. Sikora, M.C. Begelman, G.M. Madejski, et al.:
ApJ \textbf{625}, 72 (2005)

\bibitem{sol1989} H. Sol, G. Pelletier, E. Asseo: MNRAS \textbf{237},
411 (1989)

\bibitem{spruit1997} H.C. Spruit, T. Foglizzo, R. Stehle: MNRAS
\textbf{288}, 333 (1997)

\bibitem{steffen1995} W. Steffen, J.A. Zensus, T.P. Krichbaum, et al.:
A\&A \textbf{302}, 335 (1995)

\bibitem{tsinganos2002} K. Tsinganos, S. Bogovalov: MNRAS
\textbf{337}, 553 (2002)

\bibitem{unwin1997} S.C. Unwin, A.E. Wehrle, A.P. Lobanov, et al.: ApJ
\textbf{480}, 596 (1997)

\bibitem{valtonen2006} M.J. Valtonen, H.J. Lehto, A. Sillanp\"a\"a, et al.:
ApJ \textbf{643}, 9 (2006)

\bibitem{vlahakis2004} N. Vlahakis, A. Konigl: ApJ \textbf{605}, 656 (2004)

\bibitem{walker2000} R.C. Walker V. Dhawan, J.D. Romney, et al.: ApJ
\textbf{530}, 233 (2000)

\bibitem{wardle1998} J.F.C. Wardle, D.C. Homan, R. Ojha, D.H. Roberts:
Nature \textbf{395}, 457 (1998)

\bibitem{zaninetti1989} L. Zaninetti: A\&A \textbf{221}, 204 (1989)

\bibitem{zensus1997} J.A. Zensus: Ann. Rev. Astron. Astrophys. \textbf{35}, 607 (1997)

\end{thebibliography}
\end{document}